\documentclass[]{spie}  %>>> use for US letter paper
\usepackage{aas_macros} 
\usepackage{amsmath,amsfonts,amssymb}
\usepackage{graphicx}
\usepackage[colorlinks=true, allcolors=blue]{hyperref}
\usepackage{bm}
\title{Measuring the effects of General Relativity at the Galactic Center with Future Extremely Large Telescopes}

\author[a]{Tuan Do}
\author[a]{Aurelien Hees}
\author[a]{Arezu Dehghanfar}
\author[a]{Andrea Ghez}
\author[b]{Shelley Wright}
\affil[a]{UCLA Galactic Center Group, Physics and Astronomy Department,
UCLA, Los Angeles, CA 90095-1547}
\affil[b]{Center for Astrophysics \& Space Sciences, UCSD, 9500 Gilman Drive La Jolla, CA 92093 USA}

\authorinfo{E-mail: tdo@astro.ucla.edu}

\begin{document} 
\maketitle

\begin{abstract}
The Galactic center offers us a unique opportunity to test General Relativity (GR) with the orbits of stars around a supermassive black hole. Observations of these stars have been one of the great successes of adaptive optics on 8-10 m telescopes, driving the need for the highest angular resolution and astrometric precision. New tests of gravitational physics in the strong gravity regime with stellar orbits will be made possible through the leap in angular resolution and sensitivity from the next generation of extremely large ground-based telescopes. We present new simulations of specific science cases such as the detection of the GR precession of stars, the measurement of extended dark mass, and the distance to the Galactic center. We use realistic models of the adaptive optics system for TMT and the IRIS instrument to simulate these science cases. In additions, the simulations include observational issues such as the impact of source confusion on astrometry and radial velocities in the dense environment of the Galactic center. We qualitatively show how improvements in sensitivity, astrometric and spectroscopic precision, and increasing the number of stars affect the science with orbits at the Galactic center. We developed a tool to determine the constraints on physical models using a joint fit of over 100 stars that are expected to be observable with TMT. These science cases require very high astrometric precision and stability, thus they provide some of the most stringent constraints on the planned instruments and adaptive optics systems. 
\end{abstract}


\section{INTRODUCTION}
\label{sec:intro}  % \label{} allows reference to this section

The next generation of large ground-based telescopes hold the potential to open a new realm of discovery at the Galactic center, both in fundamental physics and astrophysics. The precision of the measurement of stellar orbits has  been anticipated to be powerful enough to detect very small deviations in the orbital parameters that can be used to test General Relativity and alternative theories of gravity.

While the potential for discovery has been long recognized at the Galactic center, realistic estimates the sensitivity of such experiments has not been well studied for the next generation of extremely-large ground-based telescopes (TMT, GMT, ELT). The difficulty stems from the fact that the measurement of orbits at the Galactic center requires understanding 5 major components: (1) the astrophysical content near the black hole (e.g. the predicted number of stars as well as their physical properties); (2) estimates of the performance of the adaptive optics system; (3) estimates of the performance of the science instrument and the specific configuration that will be used for observations; (4) a model for the cadence of observations and uncertainties in positional and radial velocity measurements; (5) a framework for estimating the uncertainty from fitting many orbits simultaneously to constrain physical parameters such as post-Newtonian effects on the orbits. The lack of a comprehensive study has hindered our ability to quantitatively estimate and optimize the sensitivity to tests of various physical models of gravity. 

Here, we present some of the first realistic simulations of observations of stars at the Galactic center with TMT, taking into account all 5 major components listed above. We use TMT as an example of this work because we have extensive simulations from the adaptive optics system (NFIRAOS) and integral-field spectrograph and imager (IRIS). A comparable study could be done with GMT and ELT as well. In Section \ref{sec:ao}, we present the estimates of the performance and point spread function delivered by NFIRAOS. We discuss the simulations of the instrumental performance of IRIS on Galactic center sources. Using our existing knowledge of the stellar population and distribution of stars, we predict statistically the number and types of stars that will be detected by TMT in Section \ref{sec:astrophysics}. We present our observational cadence tool and the model for the effect of source confusion used to determine the sensitivity to physical models in Section \ref{sec:models}. We discuss our resulting constraints on models such as the precession of the periaspe of the orbit of S0-2 in Section \ref{sec:results}. Section \ref{sec:conclusions} states our conclusions and prospects for future improvements. 

\section{ADAPTIVE OPTICS AND IRIS INSTRUMENT}
\label{sec:ao}

We use simulations of the AO point spread function (PSF) from the TMT NFIRAOS team in order to simulate the observations at the Galactic center. These PSFs are described in \citenum{2014AJ....147...93D} and \citenum{2013aoel.confE..83Y}. Bases on observing conditions at the Galactic center from Mauna Kea (airmass, atmospheric conditions, guide star configuration, etc.), the PSF is predicted to have Strehl ratio of about 0.7 at K-band. Since the simulations are limited to the central $1^{\prime\prime}\times1^{\prime\prime}$, spatial PSF variations should be very small so we use the same PSF for all sources. 

We use the IRIS imaging and spectroscopic simulator to produce images and spectra at the Galactic center. The properties of the simulator is described in \citenum{2014AJ....147...93D}. The simulator takes into account different noise sources as well as the throughput and sensitivity predictions for the AO system and the IRIS instrument in order to simulate realistic observations. 

\subsection{Astrometric and Radial Velocity Precision}
\label{sec:errors}
With its larger aperture and better AO correction, the astrometric precision of TMT is expected to be much improved compared to current 8-10 m telescopes. For the simulations here, we use the results of astrometric experiments performed by \citenum{2013aoel.confE..83Y}. We expect that the astrometric performance of TMT will likely be in the range of 25 to 50 $\mu$as for well measured stars, compared to about 150 $\mu$as for observations today \cite{2016ApJ...830...17B}. 

\begin{figure}[hbt]
\center
\includegraphics[width=4.0in]{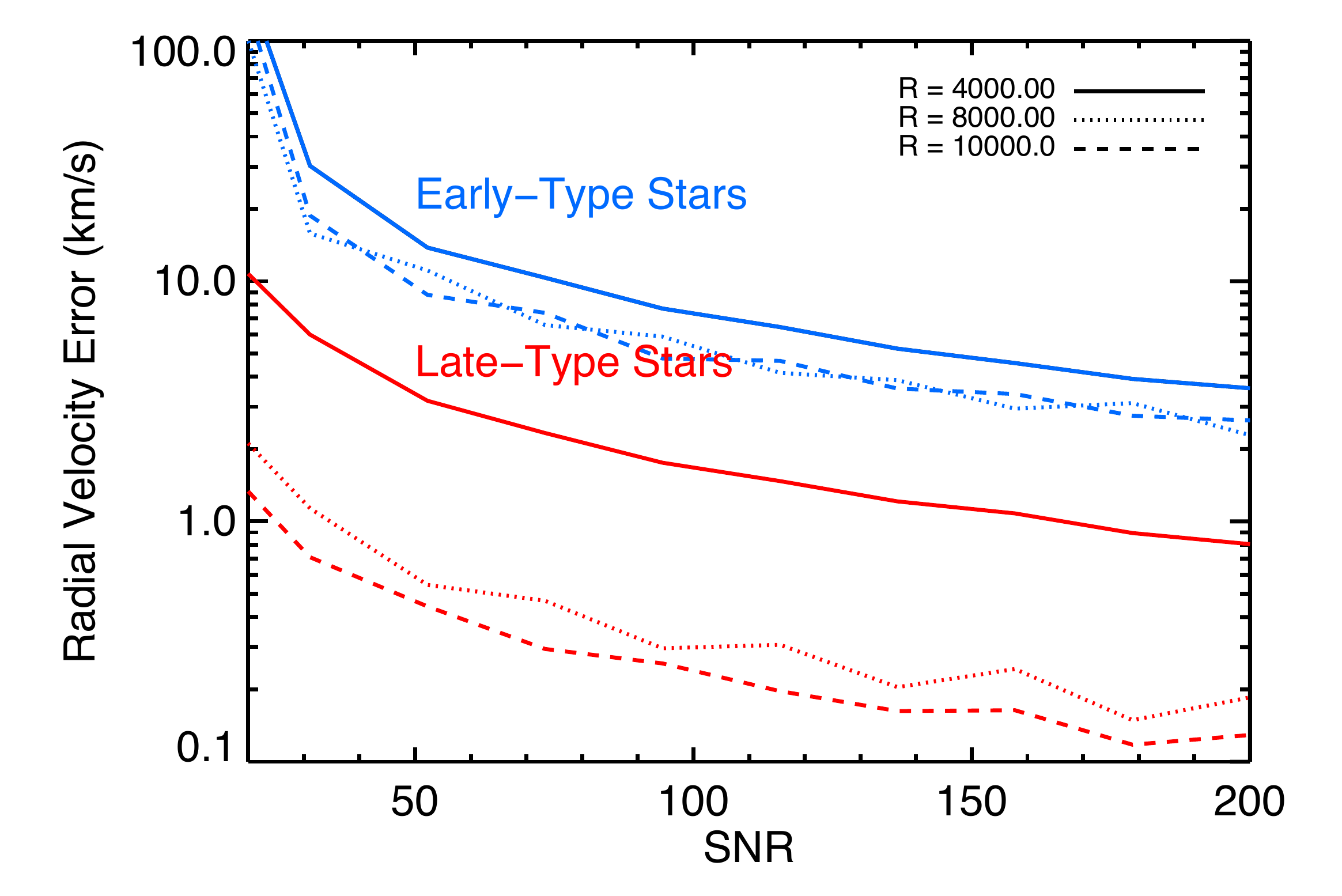}
\caption{The result of simulations showing expected radial velocity as a function of SNR at R = 4000, 8000, and 10,000 at K-band. Observations of early-type stars (blue) will have larger uncertainties compared to late-type stars (red) because they have fewer spectral lines in K-band.}
\label{fig:rv_err}
\end{figure}

The radial velocity precision is a function of both the signal to noise ratio (SNR) as well as the spectral resolution. We determine the expected radial velocity precision by simulating spectra at different SNR and at three different spectral resolution: R = 4000, 8000, and 10,000, corresponding to the potential modes for IRIS. We find that early-type stars (A-type and earlier) K-band have higher radial velocity uncertainties than late-type stars (K and M-type) because early-type stars have much fewer lines than late-type stars. At a given spectral resolution and SNR, late-type stars generally have a factor of $\sim5$ times smaller uncertainties (Fig. \ref{fig:rv_err}). We use the predictions for the radial velocity uncertainties in our orbit simulations below. One of the biggest advantage of TMT will be the increase in spectroscopic sensitivity. Today, we are spectroscopically about 50\% complete at Kp = 16.0 in this region, with the faintest sources with a measured radial velocity is about K = 17 \cite{2013ApJ...764..154D,2016ApJ...830...17B,2017ApJ...837...30G}, TMT will likely measure radial velocities down to K = 20-22. Likely, all the known short period stars without radial velocity measurements today will have measurements with TMT.

\section{WHAT LIES AROUND A SUPERMASSIVE BLACK HOLE?}
\label{sec:astrophysics}

While current 8-10 m telescopes have observed over 100 stars within just 1$^{\prime\prime}$ from supermassive black hole, future extremely large telescopes should detect many more stars -- both fainter and closer to the black hole. Currently, the detection of sources are limited by stellar confusion with both other observed stars as well as unresolved sources. This confusion limit is set by both the angular resolution of the telescope and the performance of the AO system. Typical imaging observations of stars at the Galactic center from the Keck Telescopes have Strehl ratios of 20 to 35\% with a core of about 60 mas at Kp (2.2 $\mu$m). These images represent the state of our knowledge in terms of the number and types of stars in this region. 

\begin{figure}[hbt]
\center
\includegraphics[width=6.0in]{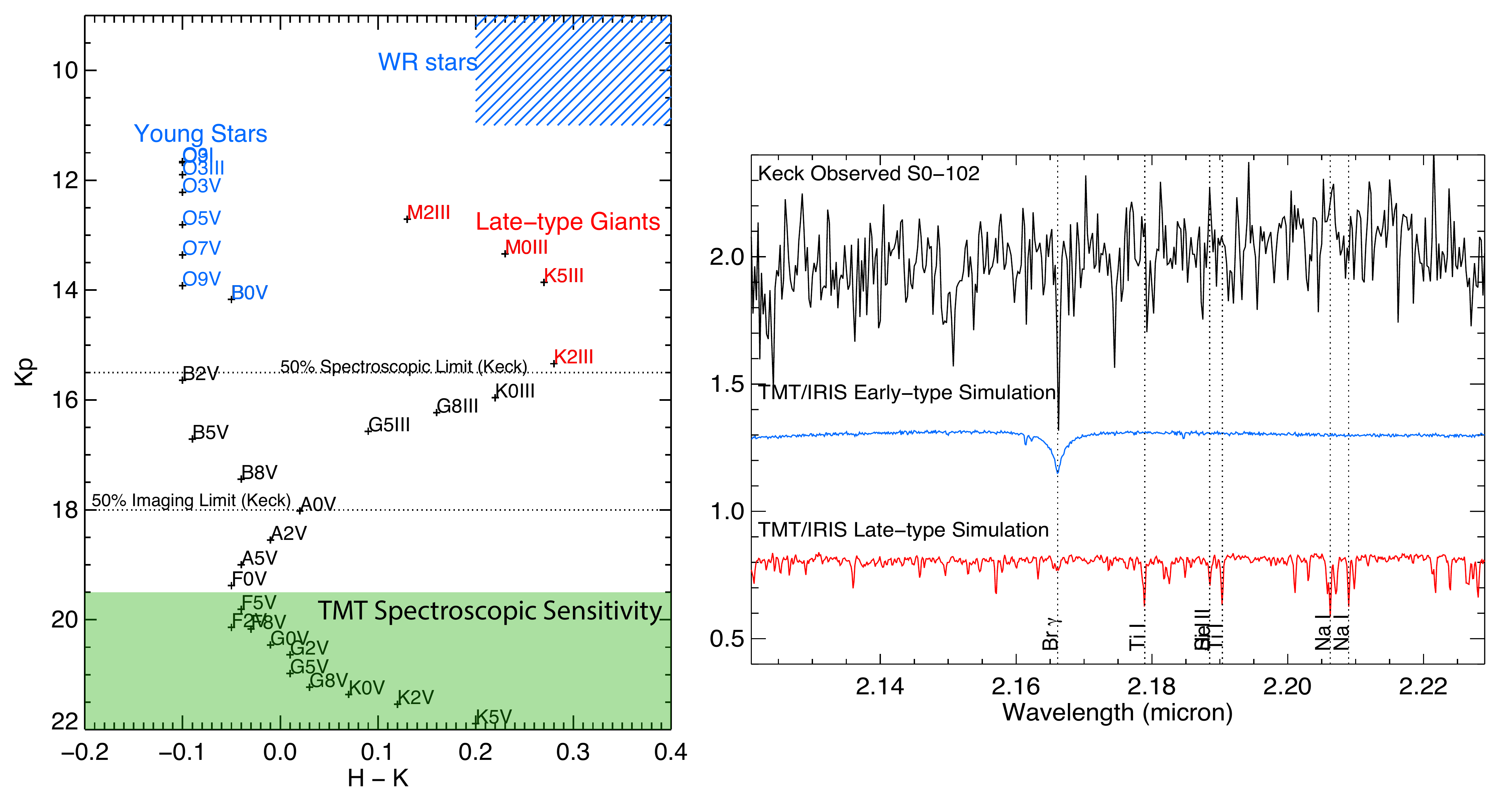}
\caption{\textbf{Left:} Theoretical color magnitude diagram of stars at the distance of the Galactic center. The H-K color is the intrinsic color of the stars. The small different in near-infrared colors and the large variations in dust extinction makes it difficult to differentiate between stars with photometry alone. Current spectroscopy from 8-10 m telescopes are limited only to the most massive young stars or red giants. TMT will provide many magnitudes of improvement in spectroscopic sensitivity. \textbf{Right:} An example of a spectrum of a short period star, S0-102 observed with Keck today. We are unable to obtain a radial velocity measurement or a spectral type of this star. With TMT, we predict that high SNR spectra will be possible for these stars. Obtaining radial velocity measurements of these stars are critical for accurate orbit determination.}
\label{fig:cmd}
\end{figure}

In order to simulate observations with TMT, we need to extrapolate our current knowledge of the stellar population and distribution of stars immediately around the black hole to simulate the properties of the underlying stars. We know today that there are two major populations of stars around the supermassive black hole: an old population from about 1 Gyr ago, and a young population formed just 4-6 Myr ago\cite{2009ApJ...703.1323D,2011ApJ...741..108P,2013ApJ...764..155L,2015ApJ...808..106S}. These populations are currently best separated spectroscopically using integral-field spectroscopy with AO, since they have about the same luminosity and color in the near-infrared (Fig. \ref{fig:cmd}). Based on the luminosity function of the currently observed young stars, \citenum{2013ApJ...764..155L} was able to fit the mass function and age of the young stellar cluster. The observations are complete only to about early B-type stars so they represent the high-mass end ($> 15 M_\odot$) of the luminosity function. Similarly, for the old stars, we can only spectroscopically observe the evolved stars which have high luminosities. Based on these stars, \citenum{2011ApJ...741..108P} derived a star formation history and mass function for the old population. Using a stellar population model, we can predict the luminosity function at fainter magnitudes for both of these populations, normalized such that the currently observed number of stars match the simulated stars at magnitudes where current observations are complete. All currently known stars, regardless of whether they have been spectroscopically identified are included in the simulations.  We simulate to a depth of K = 22 mag, about 8 magnitudes fainter than the current spectroscopic limit (Kp = 15.5) and about 4 magnitudes fainter than the imaging limit (Kp = 18.0). 

In order determine the distribution of stars in this region, we also use the currently observed density profiles of these two population to extrapolate the density of young and old stars as a function of distance from the supermassive black hole. Inherent in such an assumption is that the density profile of the unobserved sources follow the same distribution as the brighter sources. This may not be the case because of effects such as mass segregation or tidal stripping or collisions of red giants could modify the brighter population relative to the fainter one. However, since these factors are currently observationally unconstrained, we will make the simplifying assumption that the fainter population follows the brighter one. The density profile within this region of interest has the young stars steeply rising with an observed surface density profile of $\Sigma_{yng}(R) \propto R^{-1.7}$ compared to the old population with $\Sigma_{old}(R) \propto R^{-0.5}$, where R is the projected distance of the stars from the black hole. This steep density profile and the luminosity function results in more young stars than old stars within 0.04 pc of the black hole. 

The final astrophysical component is the orbital parameters for the simulated stars. Current observations of stars within the region of interest (about 0.04 pc from the black hole) show that they have orbits with orientations ($i$ - inclination, $\Omega$ - longitude of the ascending node, $\omega$ - argument of periapsis) that are isotropically distributed on the plane of the sky. This result is expected because the dynamical relaxation times for stars around the supermassive black hole should be fairly fast. The relevant time scales in this region are the timescale from General Relativistic precession and vector resonant relaxation, both of which should relax the stars within several million years \cite{2007ApJ...666..919Y}. We therefore uniformly draw the stars with randomly oriented orbital planes and from an eccentricity ($e$) distribution of the form $P(e) \propto e^{2}$ \cite{2017ApJ...837...30G}. For the semi-major axis ($a$), we draw from the radial distribution of stars discussed in the previous paragraph. We also randomly choose $T_o$, the time of closest approach of the star to the black hole. 

When combined, the astrophysical simulations of stars around the supermassive black hole contains stellar parameters such as the effective temperature, surface gravity, and age, and 6 orbital parameters ($e$, $i$, $\Omega$, $\omega$, $a$, and $T_o$). Deviations from the Keplerian orbits are included as additional free parameters, which depend on the theory of gravity being tested. 

\begin{figure}[hbt]
\center
\includegraphics[width=6.0in]{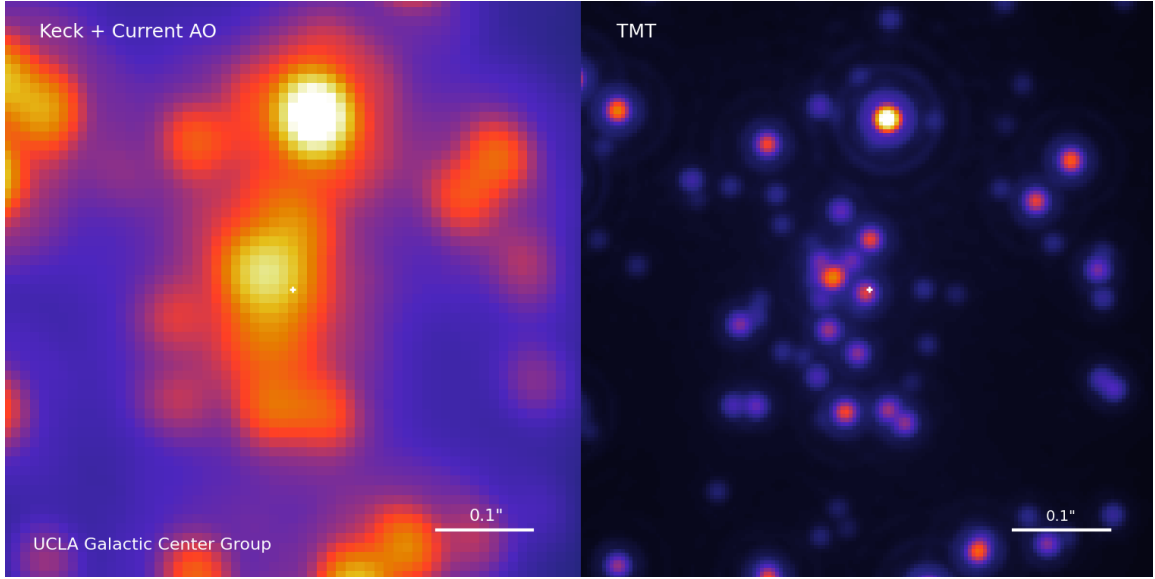}
\caption{Simulation of the central 0.5 arcsecond around the supermassive black hole Sgr A* (white cross) with current adaptive optics with the Keck Telescopes (left) compared to that with TMT (right). The greater resolving power of TMT along with better AO correction will likely allow us to detect a factor of $> 10$ times more stars in this region. We simulate how these orbit can be used simultaneously in a joint analysis to constraint tests of gravity and other science cases.}
\label{fig:keck_tmt}
\end{figure}

\begin{figure}[hbt]
\center
\includegraphics[width=5.0in]{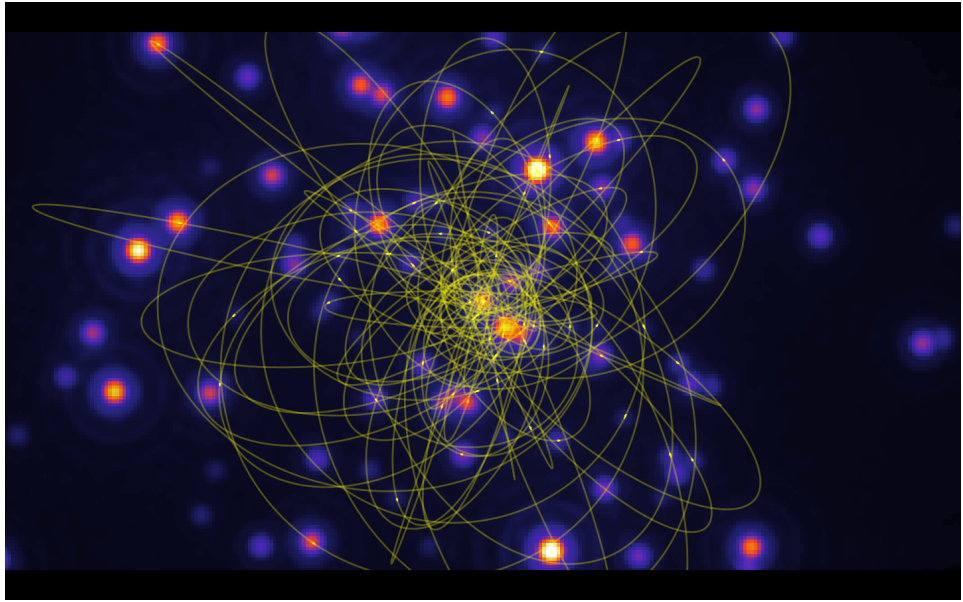}
\caption{Overlay of a sample orbits from the simulations described in Section \ref{sec:astrophysics}. TMT will likely discover stars with orbital periods below 5 years given its increased sensitivity and angular resolution compared to telescopes today.}
\label{fig:tmt_orbits}
\end{figure}

\section{ORBIT SCIENCE WITH TMT}
\label{sec:models}
We test several representative science cases that will greatly benefit from the sensitivity and angular resolution of the next generation of large ground-based telescopes: (1) measurement of the precession of the periapse predicted by General Relativity; (2) measurement of the extended dark cusp of stars and stellar remnants around the black hole; (3) measuring the black hole mass and distance to the Galactic center. 

Our goal is to perform a sensitivity analysis using realistic simulations of short-period stars in the TMT era using the following methodology:
\begin{itemize}
	\item Identify and simulate the orbital parameters of stars that will be used in our simulations.
	\item Simulate an observation plan.
	\item Model the observational astrometric and radial velocity uncertainties.
	\item Model the stellar confusion of sources and their impact on positional and radial velocity measurements. 
	\item Using all the observations epochs and related uncertainties, compute an estimate of the sensitivity to physical effects on the orbits. 
	\end{itemize}
The time baseline considered in this work starts in 1995, when high-angular resolution measurements began at Keck, up to 2047 -- after 20 years of TMT operations. We consider observations using a combination of two types of telescopes: (1) observations made using 8-10 m telescopes today  and (2) observations made using TMT--like telescope that starting in 2027. The steps of our simulations are described with more details below.

\subsection{Number of stars simulated, orbital parameters}
The sample of stars used in this simulation includes 17 well measured stars whose orbital parameters are reported by \citenum{2012Sci...338...84M,2016ApJ...830...17B,2017ApJ...837...30G} as well as 105 simulated stars that TMT is expected to detect in the central region (Section~\ref{sec:astrophysics}). All these stars are included in the source confusion analysis which is described in Section~\ref{sec:sourceconfusion}. For the purpose of the sensitivity analysis, in the Keck era, we consider the impact of combining the orbit of 17 stars in a joint orbital-analysis. In the TMT-era, we include the orbits of all 122 stars (both currently detected and simulated).

\subsection{Cadence of observations}

In order to simulate science cases, we need to model when the observations occur, the number of observations, and the observational uncertainties. As mentioned above, we consider two classes of telescopes. The first class corresponds to existing telescopes and the epochs of observation considered consist of a combination of existing observations (between 1995 and 2017) and of future simulated observations (after 2017). From 1995-2017, we use the actual observed epochs of astrometry from Keck, and observed epochs of radial velocities from Keck and VLT \cite{2016ApJ...830...17B,2017ApJ...837...30G}. From 2017-2027, we include three astrometric and three spectroscopic observations per year. 

We assume a cadence of ten astrometric and ten spectroscopic observations per year for TMT after 2027. Such a high observations frequency is required to capture the rapid motion of the fast-moving stars with short period that are expected to be discovered with the next generation of telescopes. For each year, the exact epochs of observation are randomly sampled from a uniform distribution in the observational window which spans from May 1st to September 1st. 

\subsection{Observational uncertainties}\label{sec:obs_sigma}

\subsubsection{Astrometric uncertainty}
We model the astrometric uncertainty as a function of magnitude using a constant at bright magnitudes and power-law at fainter magnitudes. This is intended to simulate the observed behavior of current astrometric errors, which flatten at bright magnitudes due to systematics such as tip-tilt jitter. At faint magnitudes, the background and photon noise dominate resulting in a power-law dependence. We approximate the astrometric errors with:
\begin{equation}\label{eq:sigma_astro}
	\sigma_\textrm{astro}(m)= A \times 10^{\alpha\left(m-m_0\right)}  \quad \textrm{for} \quad m>m_0   \qquad \textrm{and} \qquad \sigma_\textrm{astro}(m)= A \quad \textrm{for} \quad m\leq m_0 \, . 
\end{equation}
This relation depends on three constants: the constant $A$ which corresponds to the typical uncertainty reached for bright stars; the slope of the power-law $\alpha$ and the breaking point of the power-law $m_0$. These three parameters depends on the type of telescopes. The values used in this work are presented in Table~\ref{tab:sigma_astro}.
\begin{table}[htb]
\begin{center}
	\begin{tabular}{c c c c }
		\multicolumn{4}{c}{Astrometric uncertainty} \\ 
		\hline
		Telescope	&	A   [$\mu$as] &  $\alpha$	& $m_0$	 \\ \hline\hline
		Keck	    &	100	        &   0.2     &  15    \\
		TMT	        &	25          &	0.2     &  17    \\
		\hline
	\end{tabular}
\caption{Parameters characterizing the astrometric uncertainties for current and future telescopes parametrized by Eq.~(\ref{eq:sigma_astro}).}
\label{tab:sigma_astro}
\end{center}
\end{table}

\subsubsection{Radial velocity uncertainty}\label{sec:RV_sigma}
The radial velocity uncertainty depends on the spectral-type of a star (early-type or late-type) as well as its magnitude (see Section \ref{sec:errors}). We model this dependence using:
\begin{equation}\label{eq:sigma_RV}
	\sigma_\textrm{RV}(m)= B \times 10^{\beta\left(m-m_0\right)}  \quad \textrm{for} \quad m>m_0   \qquad \textrm{and} \qquad \sigma_\textrm{RV}(m)= B \quad \textrm{for} \quad m\leq m_0 \, . 
\end{equation}
The constant $B$ corresponds to the typical RV uncertainty for bright stars, the parameter $\beta$ corresponds to the slope of the power-law and $m_0$ correspond to the breaking point in the power-law. The value of these three constants depend on the spectral-type and on the characteristic of telescope. They are given in Table~\ref{tab:sigma_RV}. Typically, the next generation of extremely large telescopes is expected to improve the current RV uncertainty by one order of magnitude. 

\begin{table}[htb]
\begin{center}
	\begin{tabular}{cc c c }
		\multicolumn{4}{c}{Early-type star RV uncertainty} \\ 
		\hline
		Telescope	&	$B$   [km/s]  &  $\beta$	& $m_0$	 \\ \hline\hline
		Keck	    &	30	        &   0.17    &  14    \\
		TMT	        &	 3          &	0.17    &  14    \\
		\hline
	\end{tabular}\hspace*{2cm}
	\begin{tabular}{c c c c }
		\multicolumn{4}{c}{Late-type stars RV uncertainty} \\ 
		\hline
		Telescope	&	$B$   [km/s]  &  $\beta$  & $m_0$	 \\ \hline\hline
		Keck	    &	7	        &   0.37    &  14    \\
		TMT	        &	0.7         &	0.37    &  14    \\
		\hline
	\end{tabular}
\caption{Parameters characterizing the RV uncertainty for early-type stars (left table) and late-type stars (right table) measured with a Keck and TMT. The RV uncertainty depends on the star's magnitude $m$ through  Eq.~(\ref{eq:sigma_RV}).}
\label{tab:sigma_RV}
\end{center}
\end{table}

\subsection{Source confusion}\label{sec:sourceconfusion}
Confusion arises when a star is observed close (typically at an angular distance smaller than the angular resolution of the telescope) to another source. In such cases, the observations can be biased and in the worst case, faintest stars are completely hidden by bright stars and cannot be observed at all. The near-infrared emission from the black hole is also an important source of confusion for some stars at their closest approach to the black hole. 

The model source confusion we identified all epochs when two (or more) orbits cross. This is done by considering the Keplerian orbits for all the 122 stars used in this work. For all these confusion events, we determine if a star can be observed or if it will be hidden by the other sources nearby. We also model an increase in uncertainty from the effects of confusion. This confusion uncertainty is then added in quadrature to the observational uncertainty in Section~\ref{sec:obs_sigma}. 

We describe the model for astrometric and radial velocity confusion in more details below.

%---------------------------------astro confusion
\subsubsection{Astrometric source confusion}\label{subsec_astroconfusion}
We consider that two sources are astrometrically confused at a given epoch when their angular distance is smaller than the astrometric confusion distance limit $\delta_\text{a}$. This confusion limit depends on the typical point spread function of the detector and depends on the telescope. The values used in this work are given in Table \ref{tab_telescopes}.
\begin{table}[htb]
\begin{center}
\begin{tabular}{cc}
\hline
Telescopes	&	$\delta_\text{a}$ [mas]	\\
\hline\hline
Keck	&	63 	    \\
TMT	    &   18     \\
\hline
\end{tabular}
\caption{Astrometric confusion distance limits $\delta_\text{a}$ for Keck and TMT.}
\label{tab_telescopes}
\end{center}
\end{table}

When several stars are predicted to be observed within an astrometric radius of $\delta_\textrm{a}$, we consider only the brightest star is observed and all the fainter stars will be unobservable. In addition, we model the observed astrometric position of the brightest star to be the center of light of all the stars included within a distance $\delta_\textrm{a}$. The center of light is defined as
\begin{equation}
	x^j_{CL} = \frac{F_b x_b^j + \sum_{k\neq b} w_{kb} F_k x_k^j  }{F_b + \sum_{k\neq b} w_{kb} F_k} \, ,
\end{equation}
where the subscript $j$ denotes the two different astrometric directions, the sum of all the stars $k$ that are closer than $\delta_\textrm{a}$ with the exception of the brightest star (denoted by a subscript $b$), $x^j_k$ are the astrometric positions of the star $k$, $F_k$ is its flux and $w_{kb}$ is a weighting function to model the point spread function of the detector. In this work, we use a simple model of the core of the point spread function using a Gaussian:
\begin{align}
w_{kb} = \frac{1}{\sqrt{2 \pi} \delta_\text{a}} \exp{\left( -\frac{d_{kb}^2}{2\delta_\text{a}^2} \right)} \, ,
\end{align}
where $d_{kb}$ is the astrometric separation between the star $k$ and the brightest star $b$. 

We model the source confusion for the bright star by determining a confusion uncertainty which is given by the absolute value of the confusion bias defined as $x^j_b-x^j_{CL}$. The expression of the confusion uncertainty for the brightest star $b$ is therefore given by
\begin{equation}
	\sigma^j_\textrm{astro;conf} =  \frac{ \sum_{k\neq b} w_{kb} \frac{F_k}{F_b} \left|x_k^j-x_b^j\right|  }{1 + \sum_{k\neq b} w_{kb} \frac{F_k}{F_b}} \, ,
\end{equation}
where the subscript $j$ denotes the two different astrometric directions.

%---------------------------------RV confusion
\subsubsection{Spectroscopic source confusion}\label{subsec_rvconfusion}
If two or more sources are close to each other on the plane of the sky with approximately the same radial velocities, measuring the RV of each star with high accuracy is difficult. We have developed a methodology that identifies when this occurs; for these epochs, there are two choices: increase the uncertainty or remove the observation from epoch for that star entirely. In order to model this we consider the spectral features of the stars, classified into three categories: featureless, early-type, or late-type. Based on a star's classification, location, and brightness, we implement a series of decisions to model the effect of confusion its radial velocity measurement:
\begin{enumerate}
	\item We do not consider radial velocity measurements for the stars that are featureless in the K-band or is fainter than the spectroscopic limit of the telescope $\text{m}_\text{f}$ (see Table~\ref{tab_rv_confusion_keck}). 
	
	\item For each observation of a given star $s$, we examine nearby sources whose angular distance is closer than an astrometric distance threshold $\xi_\text{a}$ (the values used for this threshold are given in Table~\ref{tab_rv_confusion_keck}). If no neighbor is found, this observation does not suffer from source confusion.
	
	\item For all the epochs of observations when a given star has astrometric neighbor(s), we compare the magnitude of the different stars. If there is a neighboring source brighter by $\Delta m>2$, we consider the observation to be lost. On the other hand, if the star is brighter by $\Delta m>2$  than all its neighbors, we consider this measurement as unaffected by confusion. Finally, if nearby stars are within 2 magnitudes ($|\Delta m|\leq 2$) of the star $s$ under consideration, we go on with the next step.
	
	\item We identify nearby stars that are featureless. While these stars do not have spectral features, they add noise to the spectrum of the given star. We model the effect of this noise will affect the radial velocity uncertainty as: 
	\begin{align}
	\sigma_\textrm{RV;conf;featureless} =& \frac{F_\textrm{featureless}}{F_s} \ \sigma_\textrm{s,RV} \, ,
	\end{align}
	where $F_\textrm{featureless}/F_s$ denotes the flux ratio of the featureless source to the star $s$ and $\sigma_\textrm{s,RV}$ is the observational RV uncertainty of the star described in Section~\ref{sec:RV_sigma}. This uncertainty is added  to the observational uncertainty in quadrature.
	
	\item We now consider the radial velocities of nearby stars that may be early or late-type stars. If the difference in radial velocities between the nearby sources and the source under consideration is larger than the threshold, $\xi_\textrm{RV}$, given in Table~\ref{tab_rv_confusion_keck}, we do not consider the radial velocity measurement to be affected by confusion. 
		
	\item In the rare case where nearby sources have radial velocities that are not well separated from the source under consideration, we need to take into account the spectral-types of the sources involved:
	\begin{enumerate}
		\item if the star $s$ under consideration is an early-type star:
		\begin{itemize}
			\item If it is close to least one late-type star, the observation is considered missed. This is due to that the many spectral features of a late-type star will overwhelm our ability to detect the few weak lines in the early-type star. 
			\item If it is close to at least one early-type star that is brighter, the observation is considered as missed, as the spectral feature of the brighter star will dominate.
			\item  If it is surrounded by early-type stars that are all fainter, we consider an observation possible, but with additional uncertainties. The fainter stars will bias the RV measurement of the observed star according to their brightness as well as their radial velocities. We model this uncertainty as:
			\begin{equation}\label{eq:RV_conf}
				\sigma_\textrm{RV;conf}=\frac{\sum_k F_k/ F_s \left|  RV_{k}-RV_s\right| }{1 + \sum_k F_k/F_\text{s}}  \, ,
			\end{equation}
			where the sum is of the nearby early-type stars $k$, $F_k$ is the flux of the stars and $RV_k$ is the radial velocity of the stars.
 		\end{itemize}  
		\item If the star $s$ under consideration is a late-type star:
		\begin{itemize}
			\item If it is surrounded only by early-type stars, this observation is considered as successful and no additional confusion uncertainty is included.
			\item If it is surrounded by at least one late-type star that is brighter, the observation is considered as missed.
			\item If it is surrounded by at least one late-type star, with all the late-type stars in the neighborhood being fainter, the observation is considered as successful. In this case, the late-type stars that are astrometrically and spectroscopically confused with the star $s$ under consideration will impact its uncertainty. The confusion uncertainty is computed using Eq.~(\ref{eq:RV_conf}) where the sum $k$ is of the late-type stars near the star $s$.
		\end{itemize}
	\end{enumerate}
\end{enumerate}

\begin{table}[htb]
\begin{center}
\begin{tabular}{c  c  c  c  c  c  c  }
\hline
	& \multicolumn{3}{c} {Keck}  & \multicolumn{3}{c} {TMT} \\
stars			& $\text{m}_\text{f}$ 	&	$\xi_\text{a}$	[mas] &	$\xi_\text{rv}$	[km/s] & $\text{m}_\text{f}$ &	$\xi_\text{a}$ [mas]	& $\xi_\text{rv}$ [km/s]\\
\hline\hline
early-type 		& 16.0	&	80 	&	200 	& 22.0	&	26.7 		&	200 \\
late-type	    &	16.5	&	80	&	75.0 		&	22.5	&	26.7 		&	75 	\\
\hline
\end{tabular}
\caption{ The threshold magnitude $m_\text{f}$, the confusion distance limit $\xi_\text{a}$ and the radial velocity limit $\xi_\text{rv}$ for observation with both Keck and TMT telescopes.}
\label{tab_rv_confusion_keck}
\end{center}
\end{table}

\subsection{Sensitivity Analysis}
The different steps described previously allows us to derive a list of epochs for astrometric and spectroscopic observations and their related uncertainties for the 122 stars considered in this analysis. In this section, we will describe how we estimate the uncertainties on different scientific measurements using these observations in an orbital fit. 

We re-iterate that in the Keck era, only the 17 stars that are currently well measured are considered in a joint analysis. From 2027, we start to include observations from the other 105 simulated stars.

In order to estimate the uncertainties, we compute the covariance matrix related to the linearized least-square problem and to use this covariance matrix as an estimator of the uncertainty that can be reached with a given set of observations. This covariance matrix $\Sigma$ is computed as the inverse of the normal matrix $\Sigma = N^{-1}$ defined by $N= P^T \cdot P$. The matrix $P$ contains the partial derivatives of the model with respect to the parameters. More precisely  
\begin{equation}\label{eq:partial}
	P_{ij}=\frac{1}{\sigma_i}\frac{\partial M(t_i,\bm p)}{\partial p_j}\, .
\end{equation}
where $M(t,\bm p)$ represents the model to be adjusted to the observations, $\bm p$ are the parameters of the model, $t_i$ are the epochs of observations and $\sigma_i$ is the uncertainty related to the observation from the epoch $t_i$. We note that the covariance matrix obtained with this method corresponds exactly to the one that would be obtained in a Bayesian framework under the assumptions of a Gaussian likelihood, independent measurements and if the problem is sufficiently linear in the neighborhood of its minimum (in practice, if the posterior is Gaussian, the two methods will agree). Under these assumptions, the normal matrix corresponds to the Fisher information matrix. The expected uncertainty on the scientific parameters are given by the diagonal of the covariance matrix.

In this analysis, the orbital model that we use includes the Newtonian and the first post-Newtonian correction from the central SMBH and the Newtonian contribution from the extended mass. The stellar-mass density of the extended mass follows a power-law $r^{-\gamma}$, which leads to the following expression for the mass enclosed within a radius $r$
\begin{equation}
	M_\textrm{\tiny EM}(r)=M_\textrm{\tiny EM}(r_0)\left(\frac{r}{r_0}\right)^{3-\gamma} \, .
\end{equation}
The equations of motion integrated are therefore given by
\begin{equation}
	\ddot{\bm a} = - \frac{G M \bm{x}}{r^3}-G M_\textrm{EM}(r_0) \frac{r^{-\gamma}}{r_0^{3-\gamma}}\bm x +\eta \frac{G M }{c^2 r^3} \left(4\frac{G M}{r}-v^2 \right)\bm{x}  + 4 \eta\frac{G M (\bm x \cdot \bm v)}{c^2 r^3} \bm{v} \nonumber \, .
\end{equation}
The first part of this equation of motion is the standard Newtonian acceleration from the SMBH, the second part is the Newtonian acceleration due to the extended mass, the rest of this equation is the first post-Newtonian contribution from the SMBH. We included a parameter $\eta$ to allow us to parametrize a deviation from GR (i.e. in GR, $\eta=1$ whereas Newtonian gravity has $\eta = 0$). The uncertainty related to this parameter allows us to infer the signal to noise ratio (SNR) related to the detection of the GR advance of periastron ($SNR=1/\sigma_\eta$). 

The result of the integration is then transformed into astrometric and RV observations by a standard Euler rotation using the Thiennes-Innes constant. The distance to the Galactic center $R_0$ is used to transform the distance in the plane of the orbit into observed angular position in the plane of the sky. We emphasize that in this work, we include 4 parameters that are related to the astrometric position and velocity of the SMBH and one parameter that is related to the velocity of the black hole along the line of sight. We allow these parameters to be free; fixing these parameters will improve the uncertainties in the GR measurement. 

In total, the model used includes the following parameters: 6 orbital parameters per star (period, eccentricity, time of closest approach and the three regular angles), the SMBH $GM$, the distance to the Galactic Center $R_0$, the astrometric position and velocity of the SMBH ($x_0$, $y_0$, $v_{x_0}$ and $v_{y_0}$), the velocity of the SMBH along the line of sight ($v_{z_0}$), the amount of extended mass $M_\textrm{EM}(r_0)$ and eventually a parameter to characterize the test of GR: $\eta$.

\subsection{Results}\label{sec:results}
Using the procedure presented above, we run several simulations to measure the time evolution of the uncertainty related to several parameters of scientific interest. 

\begin{figure}[hbt]
\center
\includegraphics[width=6.0in]{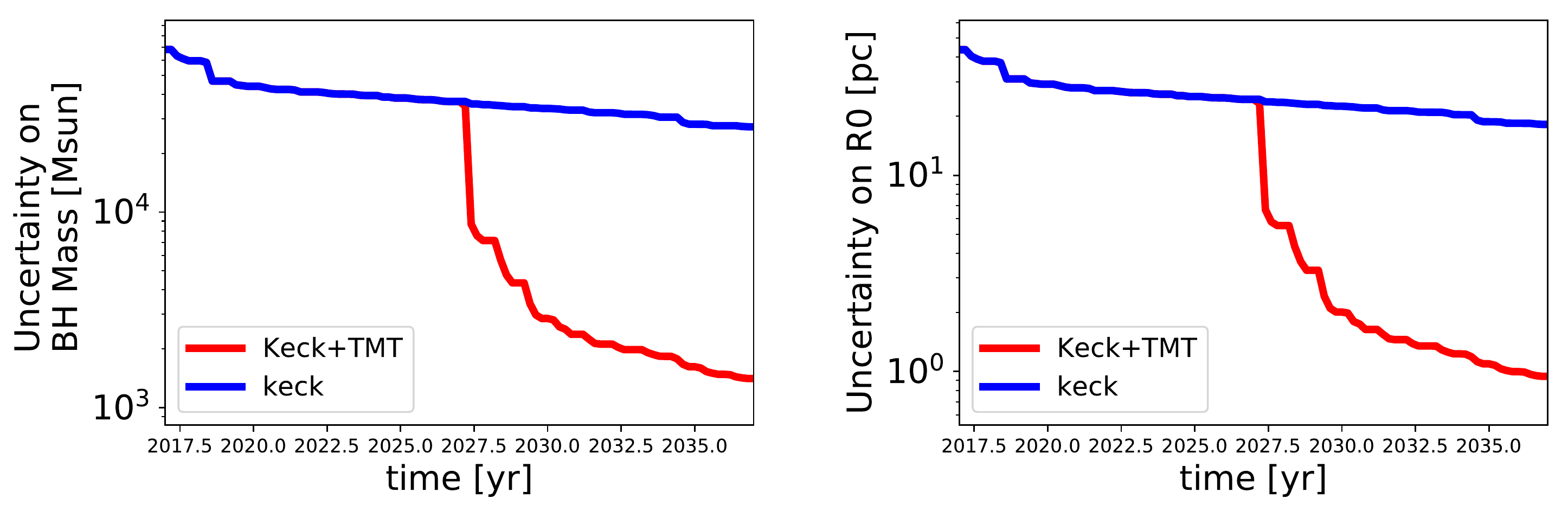}
\caption{The left panel shows the uncertainty on the black hole mass parameter in the unit of mass of the sun. The left panel displays the uncertainty on the distance of SgrA* in parsec. The red curve shows the evolution with Keck and TMT data. The blue curve on the other hand is with only Keck data.}
\label{fig:gm_r0}
\end{figure}

\begin{figure}[hbt]
\center
\includegraphics[width=6.0in]{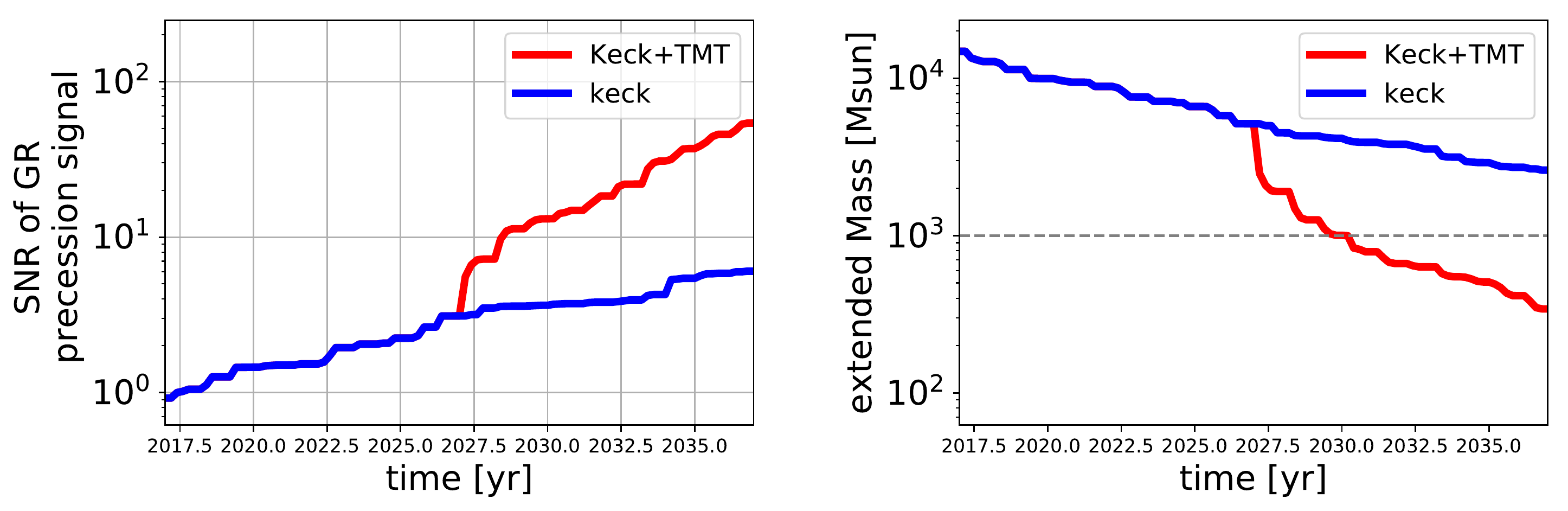}
\caption{The left panel shows the signal to noise ratio for the GR precession signal and the left panel displays the uncertainty on the extended mass parameter at the Galactic Center. The red curve shows the evolution with Keck and TMT data. The blue curve on the other hand is with only Keck data. The grey dashed horizontal line in the right panel is the extended mass value at the distance of semi--major axis of S0--2.}
\label{fig:gr_extm}
\end{figure}

The first case that we consider is a case where we assume GR to be valid (i.e. we do not consider the parameter $\eta$ in the fit). From these fits, we can see how the uncertainties on the SMBH mass, on the distance to the Galactic Center $R_0$ and on the amount of extended mass included within S0-2 apoapse evolve with time. These evolutions are presented in Figures~\ref{fig:gm_r0} and \ref{fig:gr_extm}. We consider two sets of simulations: one with no TMT observations and one where TMT observations begin in 2027. The improvement expected from TMT observations are spectacular for all parameters. More precisely, TMT observations will allow us quickly to improve the estimate on all parameters by at least one order of magnitude, reaching the impressive relative accuracy of 1 per 1000 for $GM$ and $R_0$. Regarding the extended mass, TMT allows us to probe values below $10^3$ solar masses, which will allows to distinguish between different astrophysical model\cite{2005PhR...419...65A}.

In the second simulation, we consider the parameter $\eta$ in the simulation as well. This parameter allows us to quantify the expected SNR for the relativistic advance of the periastron. This parameter can also be considered as a parameter encoding deviation from GR. As can be seen from the left panel from Figure~\ref{fig:gr_extm}, TMT will also improve the detection of relativistic effects by one order of magnitude  and allows us to further improve tests of GR in a strong-gravity regime around a SMBH\cite{2017PhRvL.118u1101H}.

In addition to the scientific cases presented here, TMT will allow us to measure the SMBH spin. A combined measurement of the SMBH spin\cite{2017arXiv170904492G} and quadrupole moment would also lead to a test of the no-hair conjecture as proposed by \citenum{2008ApJ...674L..25W}. Simulations including these effects are currently on-going.

\section{CONCLUSIONS}
\label{sec:conclusions}

The measurement of stellar orbits at the Galactic center holds the potential to push the frontiers of our understanding of the physics and astrophysics of supermassive black holes with the advent of future extremely large ground-based optical and near-infrared telescopes. While the potential tests of General Relativity and other theories of gravity at the Galactic center have been anticipated for years, this is the first study to quantitatively study the impact on different science cases using realistic models of the telescope, instrument, observing cadence, and simulations of the underly stellar cluster. We use a model of the TMT AO system, the IRIS instrument, and our current knowledge of the orbits and stellar population in our simulations. We find that in all cases, within a few years of TMT operation, there will be a large leap in the sensitivity of the orbits to physical effects such as the GR precession of the orbit and the extended mass distribution. The measurement of the precession of the periapse from General Relativity will be measured at $> 10\sigma$ significance, while fundamental parameters such as the distance to the Galactic center can be measured to better than 5 pc, which is 20 times better precision than today. The science cases outlined here are only a small sample of the potential science that is anticipated. There are also many areas that will only be discovered once this region is revealed at with unprecedented sensitivity and angular resolution.

%\acknowledgments % equivalent to \section*{ACKNOWLEDGMENTS}       
 
% References
\bibliography{/u/tdo/Documents/bibtex/prelim} % bibliography data in report.bib

\end{document}